\documentclass[twocolumn,prl,superscriptaddress,showpacs,amsmath]{revtex4}

\usepackage{graphicx}

\begin{document}

\title{Few-body bound states in dipolar gases and their detection}
\author{B. Wunsch}
\affiliation{Department of Physics, Harvard University, Cambridge MA, 02138}
\author{N.~T. Zinner}
\affiliation{Department of Physics and Astronomy, Aarhus University,  Aarhus C, DK-8000}
\affiliation{Department of Physics, Harvard University, Cambridge MA, 02138}
\author{I. B. Mekhov}
\affiliation{Department of Physics, Harvard University, Cambridge MA, 02138}
\affiliation{University of Oxford, Department of Physics, Clarendon Laboratory, Park Road, Oxford OX1 3PU, UK}
\author{S.-J. Huang}
\affiliation{Physics department and NCTS, National Tsing-Hua University, Hsinchu 300, Taiwan}
\author{D.-W. Wang}
\affiliation{Physics department and NCTS, National Tsing-Hua University, Hsinchu 300, Taiwan}
\author{E. Demler}
\affiliation{Department of Physics, Harvard University, Cambridge MA, 02138}
\date{\today}

\date{\today}
\begin{abstract}
We consider dipolar interactions between heteronuclear molecules in a low-dimensional setup consisting of two one-dimensional
tubes. We demonstrate that attra ction between molecules in different tubes can overcome intratube repulsion and complexes with
several molecules in the same tube are stable.
In situ detection schemes of the few-body complexes are proposed. We discuss extensions to many tubes and layers, 
and outline the implications on many-body physics. 
\end{abstract}
\maketitle

Few-body bound states play a crucial role in determining properties of many physical systems. 
In QCD and nuclear physics, 
quarks bind into nucleons and nucleons into nuclei. In chemistry and biology, 
chemical reactions are determined by 
properties of complexes of atoms and molecules. In soft condensed matter physics, 
self-organization of elementary 
objects into chains determines properties 
of rheological electro- and magnetofluids \cite{Klokkenburg2006}. 
In semiconducting nanostructures, like quantum wells, dots or nanotubes, 
few-body states like charged excitons and biexcitons
affect optical properties \cite{Trions}. 

A special feature of cold atom ensembles is the possibility to tune the 
two-particle interaction strength, which controls the properties of few-particle complexes.  
While most of the earlier work focused on Efimov states in systems with contact interactions \cite{Kraemer2006}, recent 
experimental progress with polar molecules \cite{exp,theory} and Rydberg atoms \cite{Rydberg} 
open interesting possibilities for studying few-particle complexes in systems 
with long-range interactions. 
These systems can provide insights into  many-body systems
with long range forces in the intriguing but poorly
understood regime of intermediate interaction strengths.
Multiparticle bound states require strong enough interactions
to form composite objects, but not too strong to avoid locking
molecules into a Wigner crystal (see e.g. \cite{Jamei2005}). Studying
dynamics  of formation of the mutiparticle composites can
help to uderstand open questions of chemical reactions in
reduced dimensions \cite{carr2009}. 

In this paper we demonstrate the stability of 
few-body states of ultracold polar molecules with long-range dipole interactions 
in a low-dimensional setup consisting of two one-dimensional tubes. This geometry 
can be produced by optical lattices or atomic 
chip traps \cite{Olattices}.
While dimers 
in bilayers of dipolar molecules have been studied before \cite{2Dbound}, the 
main result of our paper is the demonstration of the stability of few-body bound 
states with two or more molecules in the same tube. We focus on the regime where 
intratube interactions are repulsive, so that the binding 
stems from intertube attraction. We determine the stability of these complexes 
as function of the direction and the strength of the dipoles. We show the stability 
of even larger complexes and suggest a detection scheme to map out their stability 
regimes. 
%
%

\begin{figure}
\begin{center}
\includegraphics[width=\linewidth,angle=0]{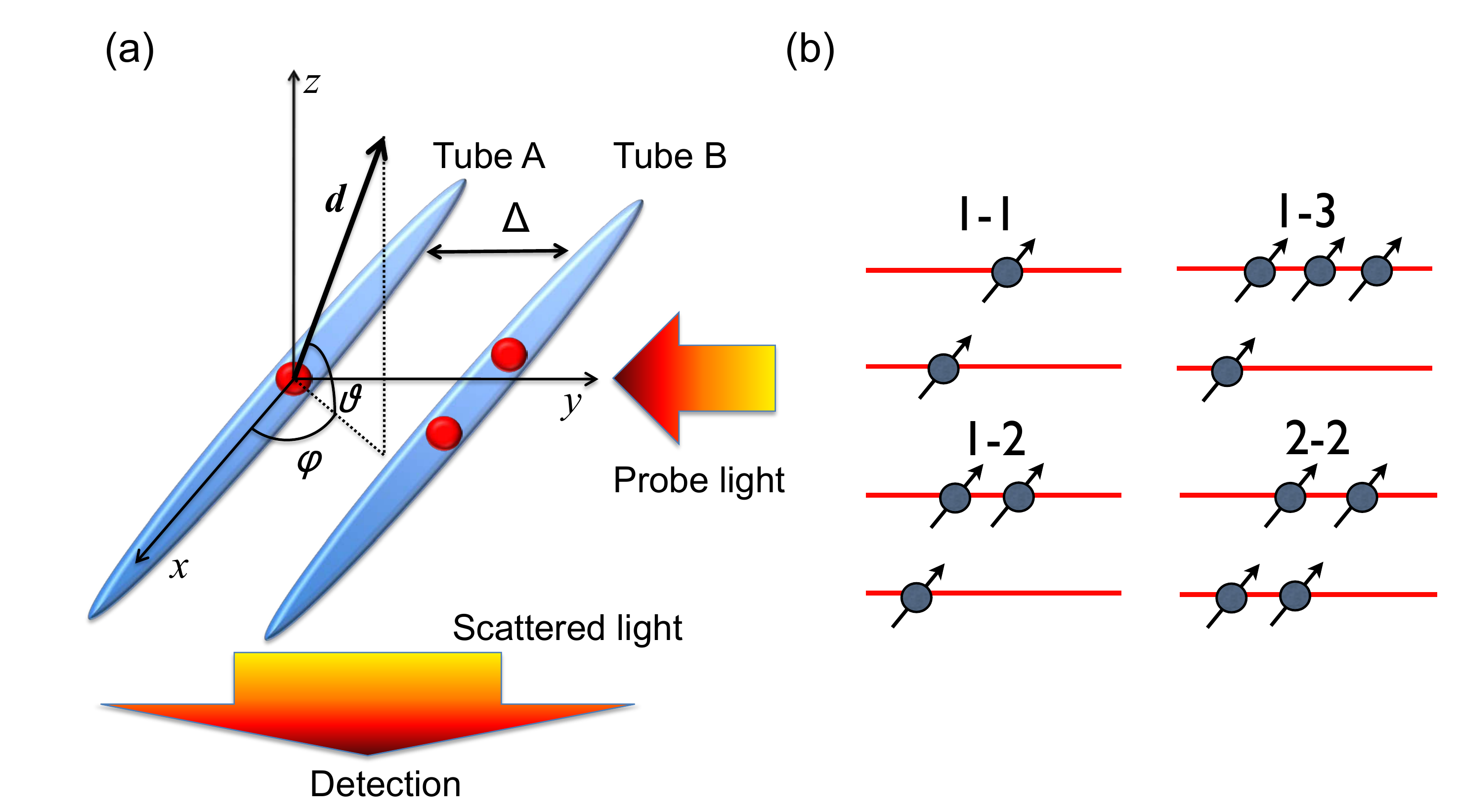}
\caption{(color online) Setup and various few-body complexes. (a) Setup. The molecules of dipole moment ${\bf d}$ move in two tubes. The probe and scattered light waves are used for the detection of complexes. (b) Notations for the complexes.}
\label{Fig:Setup}
\end{center}
\end{figure}

{\it Model.-} We consider dipolar interactions between molecules confined in tubes A and B with intertube distance $\Delta$ (Fig.~\ref{Fig:Setup}). The interaction between two dipoles aligned by an external electric field is $V_\textrm{d}({\bf r})=D^2(1-3\cos^2 \varphi_{rd})/r^3$, where $D^2=d^2/4\pi \epsilon_0$ and $\cos \varphi_{rd}={\bf r}\cdot {\bf d}/(r d)$. $\bf r$ is the relative position of two molecules and ${\bf d}$ the dipole moment. For deep 1D lattices with small transverse confinement length $l_\perp \ll \Delta$, the intertube, $V_1(x)$, and intratube, $V_0(x)$, interactions depend only on the interparticle distance along the tube direction $x$. 

The intertube interaction is $V_1(x)=D^2[1-3\cos^2{\vartheta}(\tilde{x} \cos{\varphi}+\sin{\varphi})^2/ (\tilde{x}^2+1)]/\Delta^3(\tilde{x}^2+1)^{3/2}$
The intratube interaction, $V_0(x)$, is modified at small distances by the transverse part of the wavefunction \cite{deu2010}: $V_0(x)=D^2(1-3 \cos^2 \varphi \cos^2 \vartheta)\lambda^3 f_0(\lambda \tilde{x})/\Delta^3$, 
where $f_0(u)=-u/2 +\sqrt{2 \pi} \left(1+u^2\right)\exp(u^2/2)\text{Erfc}\left(u/\sqrt{2}\right )/4$, $\lambda=\Delta/l_\perp$, and $\tilde{x}=x/\Delta$.  If the direction of the dipoles satisfies $(1-3 \cos^2 \varphi \cos^2 \vartheta)=0$, the intratube interaction vanishes. This defines the "magic angle" $\varphi_M=\arccos\left(\frac{1}{\sqrt{3}\cos(\vartheta)}\right)$. We focus on $\varphi\geq \varphi_M$, which excludes intratube attraction.
The complexes discussed here are therefore exclusively bound by intertube attraction. 
The competition of interaction to kinetic energy is determined by $U_0=m D^2/\Delta \hbar^2$. Lengths are measured in units of $\Delta$ and energies in units of $\hbar^2/m\Delta^2$.

We consider few-particle complexes of up to four molecules and denote them via the molecule numbers in the two tubes, $N_A$-$N_B$: 1-1, 1-2, 1-3, and 2-2 (Fig.~\ref{Fig:Setup}(b)). 
We use a finite difference method to obtain the eigenspectrum for the relative motion for given parameters $\vartheta,\varphi,\lambda,U_0, N_A$ and $N_B$. To reduce the size of the Hamiltonian matrices all symmetries are exploited. The stability of a complex is checked in two ways. First, we confirm that the energy of a complex is less than that of smaller few-body states causing an energy penalty for dissociation. Second, we compute the average interparticle distances, which should be finite for a bound state.

{\it Dimer.-} The simplest complex is the 1-1 dimer. For $\vartheta=0$, a dimer always profits from the attractive part of the intertube interaction and is stable for all $\varphi$ and $U_0$. However, it can be made unstable by rotating the dipoles out of the $x-y$ plane ($\vartheta \ne 0$).  The stability of the dimer depends on the integral over the intertube interaction $\int\,dx V_1(x)=2 \left(\cos^2(\varphi)\cos^2(\vartheta)-\cos(2 \vartheta)\right)D^2/\Delta^3$ and on the global minimum of the intertube interaction, $V_{1,\text{min}}$, which is negative for $\vartheta < \arccos(1/\sqrt{3})$ and positive otherwise. The directions of the dipoles can be classified into three regions according to the stability of dimers: (i) for $\int\,dx V_1(x)<0$ the  dimer is bound for any $U_0$, (ii) for $\int\,dx V_1(x)>0$ and $V_{1,\text{min}}<0$ there is a bound dimer above some critical strength $U_0$, and (iii) for $V_{1,\text{min}}\geq0$ the dimer is unbound. For the dimer it does not matter whether the molecules are fermions or bosons as long as tunneling between the layers can be neglected.

{\it Trimer and tetramers.-} Complexes with more than two molecules can be bound if the intertube attraction exceeds both intratube repulsion as well as the kinetic energy cost associated with localization. For a given $\varphi$ the intertube attraction $V_1$ is strongest for $\vartheta=0$. In order to stabilize larger complexes we will now focus on  $\vartheta=0$ and tilt the dipoles close to the magic angle [for $\vartheta=0$, $\varphi_{M}= \arccos(1/\sqrt{3})\approx 54.7^\circ$] which strongly reduces intralayer repulsion. Furthermore, we consider strong dipole interactions (large $U_0$). 

\begin{figure}
\begin{center}
\includegraphics[width=\linewidth,angle=0]{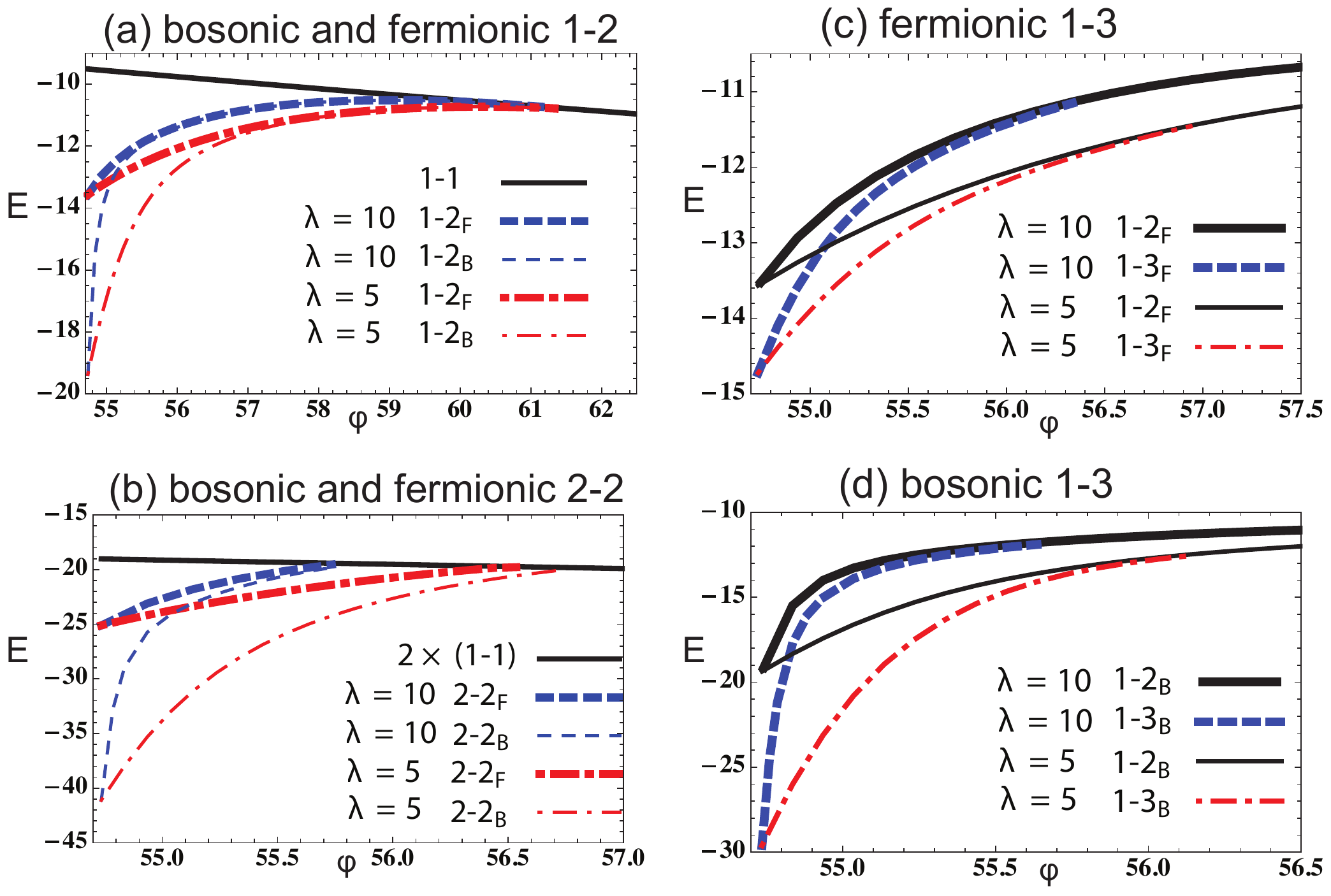}
\caption{(color online) Stability plots for (a) bosonic (subscript {\it B}) and fermionic ({\it F}) trimer 1-2, (b) bosonic and fermionic  tetramer 2-2, (c) fermionic tetramer 1-3, (d) bosonic 1-3. The complex energies $E$ as functions of the tilting angle $\varphi$, for various confinements $\lambda$, $U_0=10$. The curves for 1-1 and 1-2 (in b, c, d) are shown for comparison to demonstrate the stability of larger complexes. A complex is stable below the critical $\varphi$, where its energy is smaller than that of a smaller state. At the magic angle $\varphi_M=54.7^\circ$, the intratube interaction vanishes and the energy does not depend on the confinement $\lambda$ (curves for different $\lambda$ merge at $\varphi_M$). The energy variation is more pronounced for bosons than for fermions, and bosons are more sensitive to the confinement.}
\label{FigTr}
\end{center}
\end{figure}

Fig.~\ref{FigTr} shows that trimers and tetramer can be stable both for fermionic and bosonic molecules. Fig.~\ref{FigTr}(a) shows the range of tilting angles at which the 1-2 trimer has lower energy than the dimer (and than free particles). Shown are results both for fermionic and bosonic molecules for $U_0=10$ with transverse confinement strengths $\lambda=5$ and $10$.  We note the following common features in Fig.~\ref{FigTr}. (I) Trimer and tetramers have their energy minimum at the magic angle $\varphi_M$, since the intratube repulsion vanishes. We find that at $\varphi_M$ the energy becomes independent of the confinement, because the intertube interaction does not depend on it and intratube interaction goes to zero (at $\varphi_M$ curves for different $\lambda$ merge). At $\varphi_M$ complexes of bosonic molecules have a much lower energy than that of  fermionic molecules, since bosons can occupy the same state and profit maximally from intertube attraction, while this is forbidden for fermions by the Pauli principle.
(II) Close to $\varphi_M$ complexes of bosonic molecules have a much stronger dependence on transverse confinement and on the tilting angle  than complexes of fermionic molecules.  
This is because bosonic molecules are strongly  affected by turning on the sharp maximum of the intratube interaction at zero distance $V_0(0)=\lambda^3 \frac{(1-3 \cos^2 \varphi) D^2}{\Delta^3} \sqrt{\frac{\pi}{8}}$ as dipoles are tiltied away from $\varphi_M$. In contrast, fermions are much less affected due to the Pauli principle.
When the intratube repulsion at short interparticle distances exceeds the intertube attraction, bosons become hard-core and, generally, behave as fermions in 1D. This explains why fermionic and bosonic trimers have approximately the same energy  away from $\varphi_M$ as shown in Fig.~\ref{FigTr}(a).

\begin{figure}
\begin{center}
\includegraphics[width=\linewidth,angle=0]{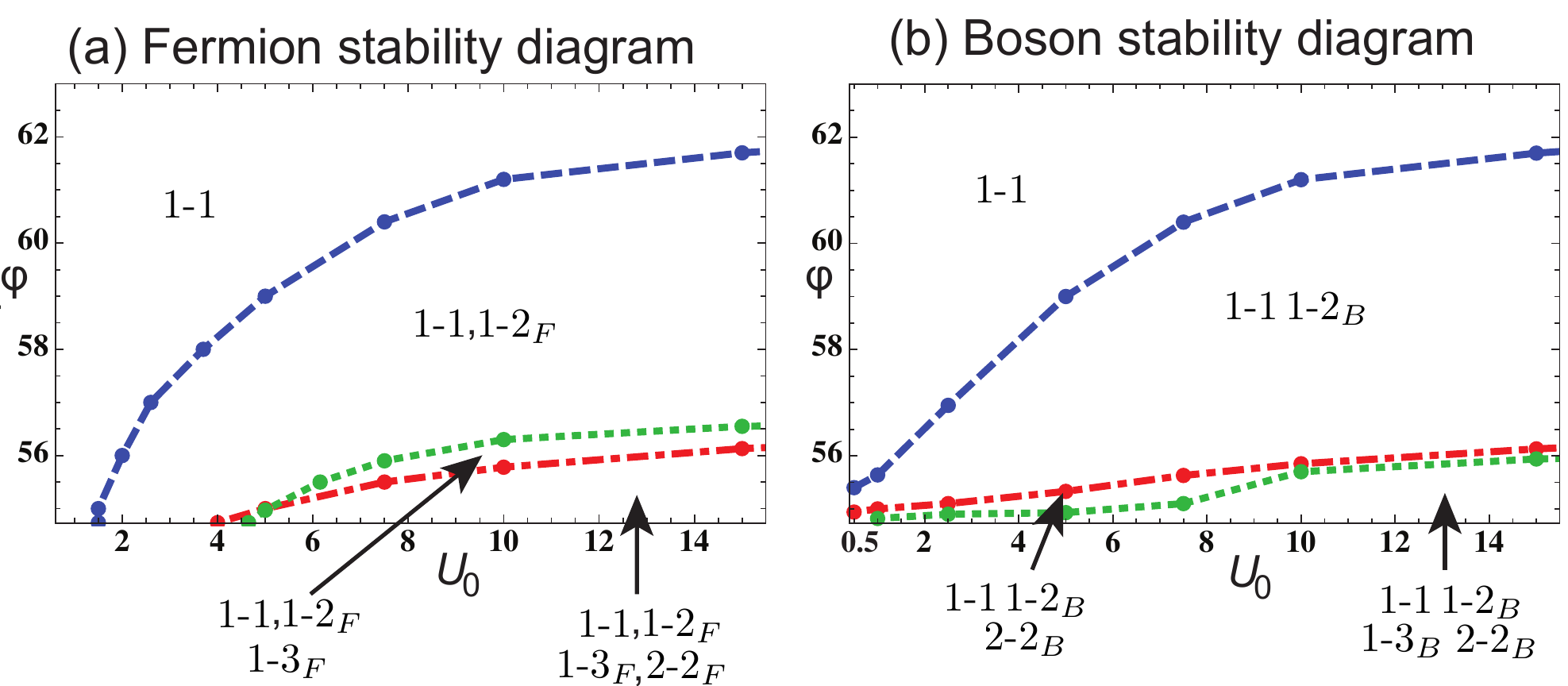}
\caption{Full stability diagram. Each region is labeled by its stable complexes. For large tilting angle, only the dimer is stable but approaching the magic angle, other complexes become stable. (a) Fermionic molecules. All states except the dimer have a critical interaction strength $U_0$. (b) Bosonic molecules. Close to the magic angle the complexes become stable for any $U_0$. Here $\lambda=10$.}
\label{FigSTAB}
\end{center}
\end{figure}

{\it Full stability diagram.-} Our results on the stability of all complexes are summarized in Fig.~\ref{FigSTAB}. Above the (blue) dashed line only the dimer is stable, while below it the trimer is also stable. The (green) dotted line marks the position, where 1-3 state becomes stable and below the (red) dashed-dotted line 2-2 state is stable. The diagram shows that with increasing number of particles in the complex the stability regime is reduced to a small interval starting from $\varphi_M$. Furthermore, by comparing the stability diagram for fermionic and bosonic molecules a major difference is evident. For fermions, there is a critical interaction strength even in the absence of intratube repulsion, while bosons do not have a critical value.

{\it Model extensions.-} Even for larger complexes we could continue the method developed here. However, already from our data we can infer the role of such states. For $N=5$, 2-3 and 1-4 configurations are possible. The stability for larger complexes in the bitube case will shrink to a small sliver near $\varphi_M$. In contrast, the trimers and tetramers are relevant over a broader range of parameters, making them more accessible in experiments. 

In a setup with planes instead of tubes, the cost of localization increases which makes complexes less stable. 
The classically preferred configurations for a bilayer are the same as for bitubes with $\vartheta=0$. Trimers or larger complexes will also be stable in bilayers, however, only for larger dipole strength. 
Dimers are always stable in the bilayer for any tilting angle \cite{2Dbound}.
In the case of more than two tubes or layers, more complex few-body states become possible. In particular, chains of molecules can form with one molecule per tube/layer \cite{wang2006}. 
By tilting the dipoles more than one molecule per layer can be bound which might lead to bifurcations of these chains. 

The few-body results discussed here have impact on the many-body problem. 
In the regime where only the dimer is stable, the fermionic many-body problem shows a BCS-BEC crossover at low temperatures \cite{BCSBEC}.  However, we have shown in Fig.~\ref{FigSTAB} that there is an extended parameter regime where not only dimers but also trimers are stable, while larger complexes are unstable.
The relevance of trimers in the many-body problem can be enhanced when the two tubes/layers have different density, as noted previously  \cite{burovski2009}.
Therefore by increasing the imbalance the system changes from a collection of interacting bosonic dimers to one of interating fermionic trimers. Trimers will modify the BCS-BEC crossover and the crystallization expected for large dipole moments \cite{crystal}.

{\it Detection.-} Multiparticle composites can be observed using several experimental techniques. For example the 1D lattice depth can be changed periodically. 
If the shaking frequency matches the binding energy of a complex it dissociates thereby heating the system. The temperature after shaking shows resonances as function of frequency \cite{shaking}. Alternatively, RF-spectroscopy can be used \cite{RF}.  
We propose to map out the stability diagram of various complexes {\it in situ}, using optical quantum nondemolition detection \cite{PRL07PRA07LasPhys09,Cirac08Eugene09Polzik09}.  Relying on coherent interference, it is sensitive to the intermolecule distances, which unambiguously reflect the complex stability.

The probe light is non-resonantly scattered by the molecules and the intensity of scattered
light is detected in the far-field (Fig.~\ref{Fig:Setup}). We propose to detect in the direction perpendicular to the tubes, where the interference condition is independent of the $x$-position of a molecule. Note that the chosen  detection angle is not the Bragg one with the largest intensity, but the diffraction minimum, where the light intensity directly reflects the molecule number fluctuations and intertube correlations \cite{PRL07PRA07LasPhys09}.

The diffraction minimum is defined such that two tubes scatter light with equal amplitudes but phase shift $\pi$. Hence, the light scattered from a tightly bound complex is zero in this direction. For different molecule numbers in two tubes, the light amplitudes at the tube positions have to be chosen differently: the amplitude ratio  $\alpha$ is 1 for 1-1 complex, 1/2 for 1-2, and 1/3 for 1-3 state. Then, the mean photon number at the diffraction minimum is $n_{\Phi}=\left|C\right|^2 \langle[N_A(W)- \alpha N_B(W)]^2\rangle$, where $N_{A,B}(W)$ are the molecule numbers in two tubes weighted by the laser profile of the finite width $W$, and $C$ is the single-molecule scattering coefficient \cite{PRL07PRA07LasPhys09}. The expectation values entering $n_{\Phi}$ are taken for the few-body ground state.

\begin{figure}
\begin{center}
\includegraphics[width=0.8\linewidth,angle=0]{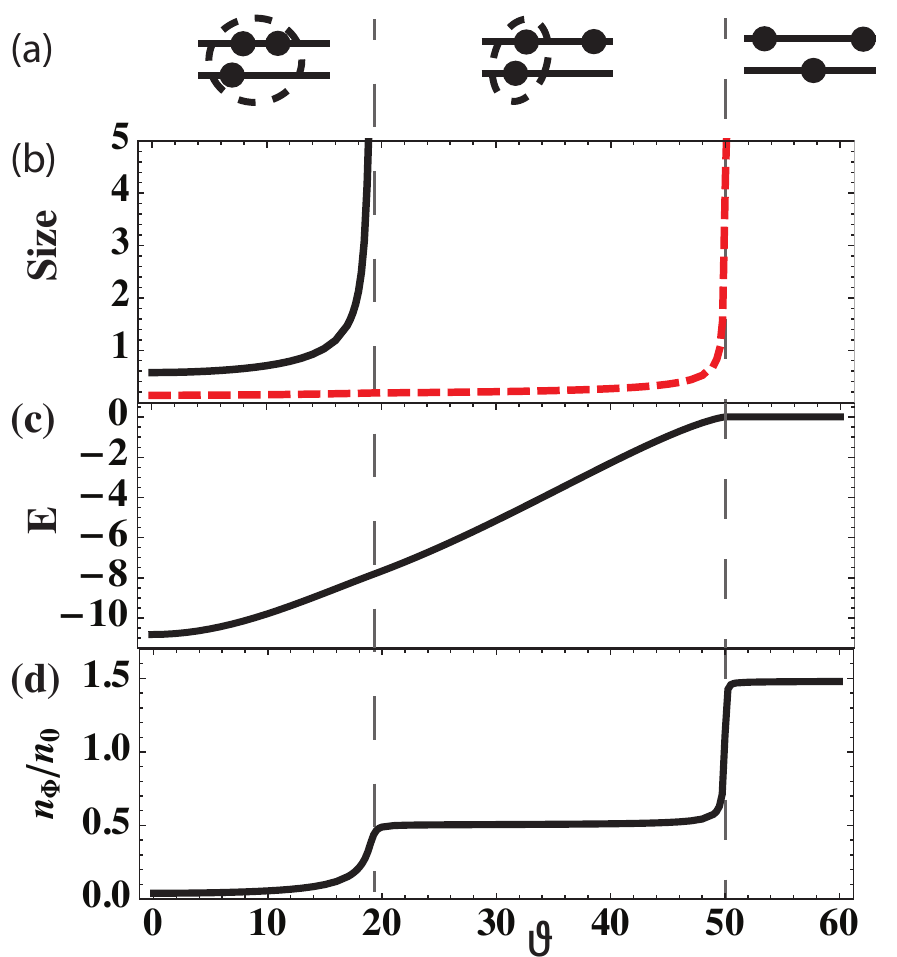}
\caption{Optical detection of trimer. (a) Three different regimes for various angle ranges. (b) Two relative distances between molecules in different tubes. The consecutive creation of 1 and 3 free molecules leads to divergence in one and then in the other distance. (c) State energy. The behavior is qualitatively different only for the last dissociation. (d) Number of photons scattered into the diffraction minimum. Each regime has its own characteristic photon numbers (0, $n_0/2$, and $3n_0/2$). The light intensity jumps show the dissociations (or creations) of a dimer and trimer. Here $U_0=10$ and $\varphi=57^\circ$, Gaussian laser profile with $W=5\Delta$.}
\label{FigDetect}
\end{center}
\end{figure}

An important property of our system is that complexes can be consecutively made stable or unstable by changing the direction or strength of the dipoles. We demonstrate that the dissociation of each complex leads to a sharp jump in the light intensity, as smaller complexes induce number fluctuations within the laser beam, which increase the intensity. As an example, the dissociation of the trimer is shown in Fig.~\ref{FigDetect}. Increasing $\vartheta$ from zero, first the trimer dissociates into a dimer and a free molecule, and then into three free molecules. At each dissociation the light intensity jumps to a new plateau with characteristic values 0, $n_0/2$, and $3n_0/2$ ($n_0$ is the photon number scattered from a single molecule). The creation of complexes correspond to light suppression.

{\it Conclusions and outlook.-}
We have shown that few-body bound states of two, three, and four dipolar molecules in a
bi-tube setup are stable over a significant range of dipole strength and direction
for both fermionic and bosonic molecules.
The complexes
are bound by the long-range intertube attraction. 
The existance of complexes can be confirmed
in a non-destructive {\it in-situ} optical detection scheme, where
the light intensity jumps precisely at the points where new bound states
form. To observe the few-body states discussed here, $U_0$ is the important parameter. We estimate that the fermionic trimer is stable for $U_0>1.5$. Such values are hard to achieve directly
in current experiments with $^{40}$K$^{87}$Rb (dipole moment $d<0.56$ Debye).
However, it may be possible to achieve the stability regime of multiparticle composites with these molecules by applying an optical lattice along the tubes. Qualitatively this may be understood as increasing the effective mass, and detailed analysis requires taking into account the tight binding dispersion.
A molecule of larger dipole moment
could also be used.
$^6$Li$^{133}$Cs has a maximum dipole moment of $d\approx5.3$ Debye \cite{exp} and  $U_0$ 
can then exceed 100 which is far within the stability regime of the complexes studied. 
For strong dipole interaction the binding energy can be large and therefore measurements 
can be performed even on non-degenerate thermal systems (for $^6$Li$^{133}$Cs and $\Delta=500$ nm the binding energy can exceed $500$nK for sufficiently large $U_0$ and close to the critical angle).
The current work proves that trimers can be stable over extended
regions of parameter space. A degenerate Fermi gas of fermonic dipolar molecules can form a collection of interacting bosonic dimers or one of interating fermionic trimers. The formation of trimers will modify the BCS-BEC crossover as well as the crystallization expected for large dipole moments. Finally, we point out that ideas presented here should also be relevant for magnetic atoms and molecules, such as Cr and Dy. However obtaining multiparticle bound states with magnetic dipolar interactions, which are much weaker than electric ones,  requires using strong in-tube optical lattice for dramatic 
suppression of the intube kinetic energy.
After completion of this work, we became aware of
the related work \cite{Dalmonte} on trimer liquids of polar
molecules in coupled tubes.

Special thanks to D. Pekker, J. von Stecher, M. Lukin, and S. Rittenhouse.  We appreciate funding by DFG
grant WU 609/1-1, FWF project J3005-N16, and EPSRC Project EP/I004394/1 and support from the Army Research Office via DARPA OLEprogram, CUA, NSF Grant No. DMR-07-05472, AFOSR Quantum Simulation MURI, AFOSR MURI on Ultracold Molecules, and the ARO-MURI on Atomtronics.

\end{document}